\documentclass[twocolumn,showpacs,superscriptaddress,preprintnumbers,amsmath,amssymb,aps,prb]{revtex4}
\usepackage{graphicx}
\usepackage{mathptmx}
\usepackage{verbatim}
\usepackage{graphicx}
\usepackage{amsmath}
\usepackage{amssymb}
\usepackage{amsfonts}
\usepackage{mathrsfs}
\usepackage{amsmath} 
\usepackage{color}
\usepackage{graphicx}
\usepackage{bm} 
\usepackage{amssymb}
\usepackage{xspace}
\usepackage{hyperref}
\usepackage{float}
\usepackage{bbold}
\usepackage{tabu}
\usepackage{textcomp}
\usepackage{dcolumn}

\newcolumntype{L}[1]{>{\raggedright\arraybackslash}m{#1}}
\newcolumntype{C}[1]{>{\centering\arraybackslash}m{#1}}
\newcolumntype{R}[1]{>{\raggedleft\arraybackslash}m{#1}}
\newcolumntype{N}{@{}m{0pt}@{}}
\begin{document}
\title{Topological phases in {\it N-}layer {\it ABC-}graphene boron-nitride moire superlattices}
\author{David Andr\'es Galeano Gonz\'alez}
\affiliation{Instituto de F\'isica, Universidad de Antioquia, AA 1226, Medell\'in, Colombia}
\affiliation{Department of Physics, University of Seoul, Seoul 02504, Korea}
\author{Bheema Lingam Chittari}
\affiliation{Department of Physical Sciences, Indian Institute of Science Education and Research Kolkata, Mohanpur 741246, West Bengal, India}
\author{Youngju Park}
\affiliation{Department of Physics, University of Seoul, Seoul 02504, Korea}
\author{Jin-Hua Sun}
\affiliation{The Research Institute of Advanced Technologies, Ningbo University, Zhejiang 315211, P. R. China}
\author{Jeil Jung}
\email{jeiljung@uos.ac.kr}
\affiliation{Department of Physics, University of Seoul, Seoul 02504, Korea}
\affiliation{Department of Smart Cities, University of Seoul, Seoul 02504, Korea}
\begin{abstract}
Rhombohedral $N=3$ trilayer graphene on hexagonal boron nitride (TLG/BN) hosts gate-tunable, valley-contrasting, nearly flat topological bands that can trigger spontaneous quantum Hall phases under appropriate conditions of the valley and spin polarization. 
Recent experiments have shown signatures of {\it C} = 2 valley Chern bands at 1/4 hole filling, in contrast to the predicted value of {\it C} = 3. We discuss the low-energy model for rhombohedral {\it N-}layer graphene  ({\it N} = 1, 2, 3) aligned with hexagonal boron nitride (hBN) subject to off-diagonal moire vector potential terms that can alter the valley Chern numbers.  
Our analysis suggests that topological phase transitions of the flat bands can be triggered by 
pseudomagnetic vector field potentials associated to moire strain patterns, and that a nematic order with 
broken rotational symmetry can lead to valley Chern numbers that are in agreement with recent Hall conductivity observations.
\end{abstract}
\maketitle
\section{Introduction}
In recent years, magic-angle twisted bilayer graphene~\cite{bistritzer,super1,mott1} (tBG) has emerged as a platform for exploring correlated insulating phases and unconventional superconductivity in compositionally simple systems, owing to the possibility of achieving extremely narrow bandwidths where Coulomb repulsion energies can dominate the kinetic energy of electrons.~\cite{bistritzer, dillonwong, super1, mott1, super2, super3, Young_quantized, zaletel, goldhaber-gordon, mott2, nadj, daixi, stephen2018, Chittari_2018, leconte2019relaxation, wei2019, Fidrysiak, vafek, koshino2, vishwanath,vishwanath2, vishwanath3, lee2019, chichinadze2019nematic, shi2019largearea, PhysRevB.100.155421, ma2019discovery, hesp2019collective, fern2019nematicity, stepanov2019interplay, zhang2019topological, Bernevig_PRL}
The pool of moire materials exhibiting such behaviors has rapidly expanded beyond twisted bilayer graphene to include
trilayer graphene on hexagonal boron nitride (TLG/BN),
~\cite{chittari, Nat_TLGBN, NatPhys_TLGBN, fengwang2019, yankowitznature, PhysRevB.100.195413, PhysRevLett.122.146802, zhang2019spin, nanolettKIM, 2019ferromagnetism, PhysRevB.99.205150, komsan2019, ulloa2012, bokdam, sanjose3, Wallbank2013, ZHU20181087} 
or double bilayer graphene~(tDBG)~\cite{tbbg, tbbg1, tbbg2, tbbg3, tbbg4, tbbg5, tbbg6,senthil}
as representative systems where the perpendicular electric field can control the flatness of the low-energy moire bands and achieve the narrow bandwidths over a wider range of the twist angles, without requiring high-precision rotation as in tBG. 
Studies have evaluated the feasibility of engineering moire flat bands in gapped Dirac materials,~\cite{srivani2019,song2019}
suggesting TMDC bilayers~\cite{feenstra, 2dheterojunctionTMD1, 2dheterojunctionTMD2, 2dheterojunctionTMD3,  song_NanNaTech}
as platforms for identifying nearly flat bands where one can benefit from the aforementioned looser constraints on twist angle 
precision.~\cite{srivani2019, mitmanish, fengcheng, fengcheng2019, tmdctwist1, tmdctwist2, bi2019excitonic, zhang2019moir, wangNSR, sushko2019high, tang2019wse2ws2,FengWang_TMD, an2019interaction} 

The finite valley Chern number of the moire bands underlie the anomalous Hall effects observed in 
transport experiments when the degeneracy of the bands are lifted via Coulomb interactions.~\cite{senthil,chittari2019}
Charge Hall conductivity signals were observed in twisted bilayer graphene nearly aligned with hexagonal boron nitride (tBG/BN)
at 3/4 filling densities,~\cite{goldhaber-gordon,Young_quantized} and closely related traces of quantum anomalous Hall effects were observed in TLG/BN.~\cite{Nat_TLGBN}
Contrary to the expectations of a charge Hall conductivity of $\sigma_{xy} = 3 e^2/ h$, which is consistent with the predicted $K$ valley Chern number $C = 3$ of hole bands for TLG/BN,~\cite{chittari2019} the experiments showed a quantized anomalous Hall conductivity of $\sigma_{xy} = 2 e^2/ h$, which is consistent with $C = 2$.~\cite{Nat_TLGBN}
In this paper we explore the valley Chern number phase diagram of G/BN, BG/BN, and ABC stacked TLG/BN structures, that can be described using the low-energy $N$-chiral Dirac model subject to moire patterns, 
in an attempt to identify the system parameter conditions that can alter the valley Chern numbers. The manuscript is structured as follows.
In Sec.~II we introduce the model Hamiltonian, 
in Sec.~III we present the electronic structure results comprising 
the valley Chern number phase diagrams,
and lastly in Sec.~IV we present the summary and discussions.
\section{Model Hamiltonian}\label{sec:Hamilton}
The low-energy model Hamiltonian of rhombohedral $N$-layer graphene on hexagonal boron nitride at the $K$-valley
subject to substrate moire patterns $H_{\xi}^{M}$ is given by
\cite{chittari}
\begin{equation}\label{eq:H_N}
H_N^{\nu,\xi}=\frac{\upsilon_0^N}{(-t_1)^{N-1}}\begin{bmatrix}
0 & \left( \pi^\dagger \right)^N\\
\pi^N & 0
\end{bmatrix} + \Delta\sigma_z +  H^{\rm R}_{N} + H_{\xi}^{M}
\end{equation}
\noindent where $\xi=\pm 1$ distinguishes the two possible (0$^{\circ}$ and 60$^{\circ}$) alignments between the layers of graphene and BN. 
The first term in equation (\ref{eq:H_N}) describes the low-energy {\it N-}layer graphene $2 \times 2$ 
Hamiltonian containing the momentum operator $\pi=\nu p_x+i p_y$, 
where $\nu=1$ is used for the principal valley $K$ of graphene, unless stated otherwise. 
The second term of equation (\ref{eq:H_N}) introduces the interlayer potential difference proportional to the mass term through $\Delta$. 
The third term, $H^{\rm R}_{N}$, describes the remote hopping term corrections for the {\it N-}layers. 
We model the remote hopping term corrections for TLG with $N=3$,~\cite{jjung_unpublished} as shown below:
\begin{eqnarray}
\label{eq:H_N3}
H^{\rm  R}_{3, \nu} &=&
\left[ 
\left(
\frac{2 \upsilon_0 \upsilon_3 \pi^2}{t_1} + t_2
\right)   \sigma_{x} 
 \right]
\\
&+&
 \left[  \frac{2 \upsilon_0 \upsilon_4 \pi^2}{t_1}  - \Delta'  + \Delta'' \left(
1- \frac{3\upsilon_0^2\pi^2}{t_1^{ 2}}
\right)  \right]  \mathbb{1}.
\nonumber  
\end{eqnarray}
The effective hopping parameters for rhombohedral trilayer graphene that 
fits the local density approximation (LDA) bands are $t_0=-2.62$~eV, $t_1= 0.358$~eV, $t_2=-0.0083$~eV, $t_3=0.293$~eV, and $t_4=-0.144$~eV, 
where associated velocities are defined as $\upsilon_m = \sqrt{3} a \, | t_m |  / 2\hbar$, 
using $a=2.46\,\AA$ as the lattice constant of graphene, and the constants in the diagonal terms
are $\Delta' = 0.0122$~eV and $\Delta'' = 0.0095$~eV.

In the case of $N = 2$, the parameters $t_2$ and $ \Delta''$ drop out, and $ \Delta' = 0.015$~eV is used for bands 
obtained within the LDA~\cite{jjung_bilayer}
and the remote hopping terms are captured as
\begin{eqnarray}
\label{eq:H_N2}
H^{\rm  R}_{2, \nu} &=&
-\upsilon_3   \left(\begin{matrix}
0 & \pi\\
\pi^\dagger & 0
\end{matrix} \right)
\nonumber 
+
\frac{ \upsilon_3}{4\sqrt{3}} \left( \begin{matrix}
0 & \pi^{\dagger 2}\\
\pi^2 & 0
\end{matrix}\right)\\
\nonumber 
&+&
 \frac{ \upsilon_0}{t_1} \left[\frac{ \Delta^{\prime} \upsilon_0} {t_1} +2\upsilon_4 \right]\left(\begin{matrix}
\pi ^\dagger \pi & 0\\
 0 & \pi \pi^\dagger
\end{matrix}\right).
\nonumber 
\end{eqnarray}

In the case of $N = 1$, all the remote hopping terms drop out and $H^{\rm  R}_{1, \nu} = 0$.
\par The last term of equation (\ref{eq:H_N}) 
\begin{equation}
H_{\xi}^{M} = H_{\xi}^{V}  + H_{\xi}^{A}
\end{equation} 
is the effective moire potential term induced by the hBN layer that consists of the diagonal $H_{\xi}^{V}$ and off-diagonal
$H_{\xi}^{A}$ terms.
We use the local commensurate stacking vector $\vec{d} = (d_x,d_y)$ between the substrate and the contacting graphene layer,
where the stacking vector $\vec{d}$ and the real space position $\vec{r}$ are related through
\begin{eqnarray}
\vec{d}(\vec{r}) \simeq \varepsilon \vec{r}  +  \theta \hat{z}\times \vec{r}
\end{eqnarray}
in the small angle approximation.~\cite{jung2014} 
Here, $\theta$ is the relative twist and $\epsilon = (a - a_{BN})/a_{BN}$ is the lattice constant mismatch between the graphene and hBN layers, where $a$ and $a_{BN}$ are the lattice constants of graphene and hBN layers respectively.

The diagonal term of the Hamiltonian in real space is given by 
\begin{eqnarray}
H_{\xi}^{V} (\vec{r})  =  V_{AA/BB}^M(\vec{r})\left[\frac{\mathbb{1}+\xi\sigma_z }{2}\right]
\label{diagterm}
\end{eqnarray}
where the moire potential function is given by 
\begin{eqnarray}
V_{AA/BB}^M(\vec{r}) = 2C_{AA/BB}{\rm Re}\left[ e^{i\phi_{AA/BB}} f^{\xi}(\vec{r}) \right],
\label{moirepot}
\end{eqnarray}
which in turn depends on the auxiliary function
\begin{eqnarray}
f^{\xi}(\vec{r}) = \sum_{m=1}^{6}e^{i\xi{ {\tilde G_m}}\cdot{\vec r}}\frac{ \left( 1+(-1)^m \right)}{2} 
\end{eqnarray}
expressed using six moire reciprocal lattices $\tilde G_{m=1..6} = \hat R_{2\pi(m-1)/3} {\tilde G_1}$ 
successively rotated by 60$^{\circ}$. 
The moire reciprocal lattice vector ${\tilde G_1} \approx \epsilon \vec G_1-\theta \vec z \times \vec G_1$ 
is related to the following reciprocal lattice vector of graphene $\vec G_1 = [0, 4\pi/(\sqrt{3}a)] $. 
For the diagonal terms of the moire potentials we use the parametrization of a G/BN interface~\cite{jung2014} projected onto only one of the sublattices.~\cite{chittari}
The moire potential in Eq.~(\ref{moirepot}) can be broken down in terms of inversion 
symmetric and asymmetric patterns, see Refs.~[\onlinecite{Wallbank2013,jung2017}]
and appendix A for more details. 

\par The intersublattice off-diagonal term $H_{\xi}^{A} (\vec{r})$ introduces coherence between the low energy sublattices in the system and could 
arise in multilayer graphene with BN interfaced systems due to higher order perturbation corrections that couple the diagonal moire pattern terms with interlayer tunngling. 
This term can be modeled as 
\begin{eqnarray}
H_{\xi}^{A} (\vec{r}) = \vec{A}^\xi(\vec{r})\cdot \sigma^{\xi}_{xy}  
\label{vecpot}
\end{eqnarray}
where $\sigma^{\xi}_{xy} = (\sigma_x, \xi\sigma_y)$ is the Pauli matrix vector and the 
pseudomagnetic vector potential that we model through
\begin{eqnarray}
\vec{A}^\xi(\vec{r}) = V^M_{AB} \, \vec{ \mathbb{\nabla} }_{\vec{r}} \, {\rm Re} \left[ e^{i\phi_{AB}} f^{\xi}(\vec{r}) \right],
\end{eqnarray}
where the prefactor 
\begin{eqnarray}
V^M_{AB} = 2C_{AB}\left[\cos(\tilde\theta) \hat{z}\times \frac{\mathbb{1}}{|\tilde G|} - \sin(\tilde\theta)\frac{\mathbb{1}}{|\tilde G|}\right]
\end{eqnarray}
depends on twist angle and lattice constant mismatch
\begin{eqnarray*}
\cos({\tilde\theta}) = \frac{\alpha\cos(\theta)-1}{\beta}; ~~
\sin(\tilde\theta) = \frac{\alpha\sin(\theta)}{\beta}\\ \nonumber
\beta = \sqrt{\alpha^2-2\alpha\cos(\theta)+1}; ~~\alpha = 1+\epsilon.
\end{eqnarray*}
We will show that this off-diagonal vector potential $\vec{A}^{\xi}(\vec{r})$ in Eq.~(\ref{vecpot}) can alter the 
Chern number phase diagram by breaking the rotational symmetry and 
modify the valley Chern numbers depending on system parameter values.

In a previous report~\cite{chittari} on the electronic structure of TLG/BN the moire potentials were modeled 
to act at the low energy sublattice of the contacting graphene layer and zero direct interlayer coherence between the
low energy sites of the top and bottom layers was assumed by using $C_{AB} = 0$ zero off-diagonal term.
There the $C = N$ proportionality to layer number was verified to up to three layers
for either valence or conduction bands depending on the sign of the electric field and hBN substrate alignment orientation. 
As a matter of convention, in this work we assume that the $C$ valley Chern number is associated to the $K$ valley of the multilayer graphene layer while a time reversal symmetric counterpart is assumed for the $K'$ counterpart. 
The zero or finite integer value of the valley Chern number in each band is attributed to the sum of the primary and secondary Chern weights typically concentrated near the gap opening points in the mBZ leading to the total valley Chern number, such that $C^{e/h} = w^{e/h}_P + w^{e/h}_S$. While the primary Chern weight near charge neutrality 
is set by the interlayer potential difference, the secondary Chern weights near the mBZ boundaries depend on the moire potential parameters that generate the avoided secondary gaps at the mBZ corners. 
We will show in our analysis that the secondary Chern weights are easily altered based on the choice of the off-diagonal moire potential terms.  
\begin{figure}[t]
\includegraphics[width=8cm,angle=0]{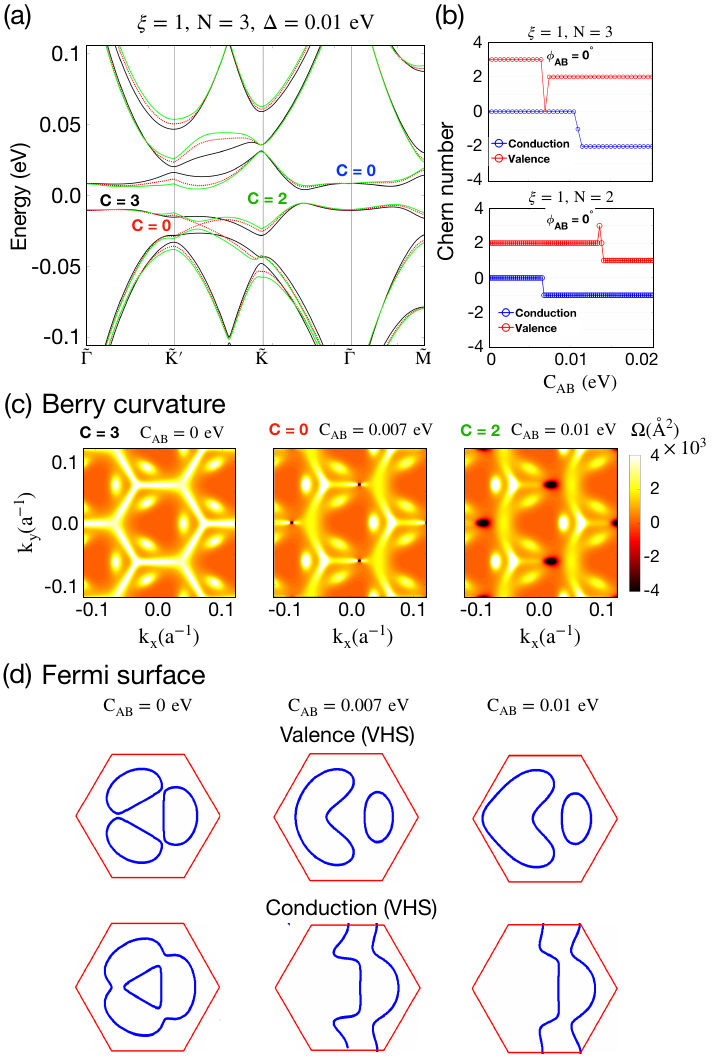} 
\caption{(color online) 
Band structures, valley Chern numbers, Berry curvatures and the Fermi surfaces at the vHS of the low energy bands in 
TLG/BN when a finite off-diagonal term as in Eq.~(\ref{vecpot}) is added to the Hamiltonian. 
(a) Band structures for $\xi = 1$, $N=3$ with $\Delta = 0.01$~eV, for different values of the off-diagonal 
pseudomagnetic field patterns proportional to $C_{AB} = 0 $ (black), $C_{AB} = 0.007 $~eV (red), 
and $C_{AB} =  0.01 $~eV (green). For all three cases we use the $\phi_{AB} = 0$ phase term. 
The respective $K$ valley Chern numbers for the valence bands are for $C_{AB} = 0$,  
$C = 3$ (black), for $C_{AB} = 0.007$~eV, $C = 0$ (red) and for $C_{AB} = 0.01 $~eV,  $C = 2$ (green). 
The valley Chern number of the conduction bands are in all cases $C = 0$. 
(b) Topological phase transitions as a function $C_{AB}$, keeping $\phi_{AB} = 0$ for the phase term when $N = 3$ and  $N = 2$ for $\xi = 1$ and $\Delta = 0.01$~eV.
We note that for $N = 3$ there is a $C = 3 \rightarrow 2$ transition for the valence bands and from $C =0 \rightarrow -2$ for the conduction bands. 
For $N = 2$ there is a $C = 2 \rightarrow 1$ transition in the valence bands and from $C = 0 \rightarrow -1$ for the conduction bands.
(c) The Berry curvatures for the three valence bands giving the different Chern numbers. 
The closure of the band gap associated with the topological phase transitions
can be identified to take place between $\tilde K^{\prime}$ and $\tilde K$. 
(d) Fermi surface contours at the vHS for the low energy valence and conduction bands where we can observe breaking of triangular rotational symmetry when $C_{\rm AB} \neq 0$.
}
\label{elect}
\end{figure}
\begin{figure}[t]
\label{schematic}
\includegraphics[width=8.5cm,angle=0]{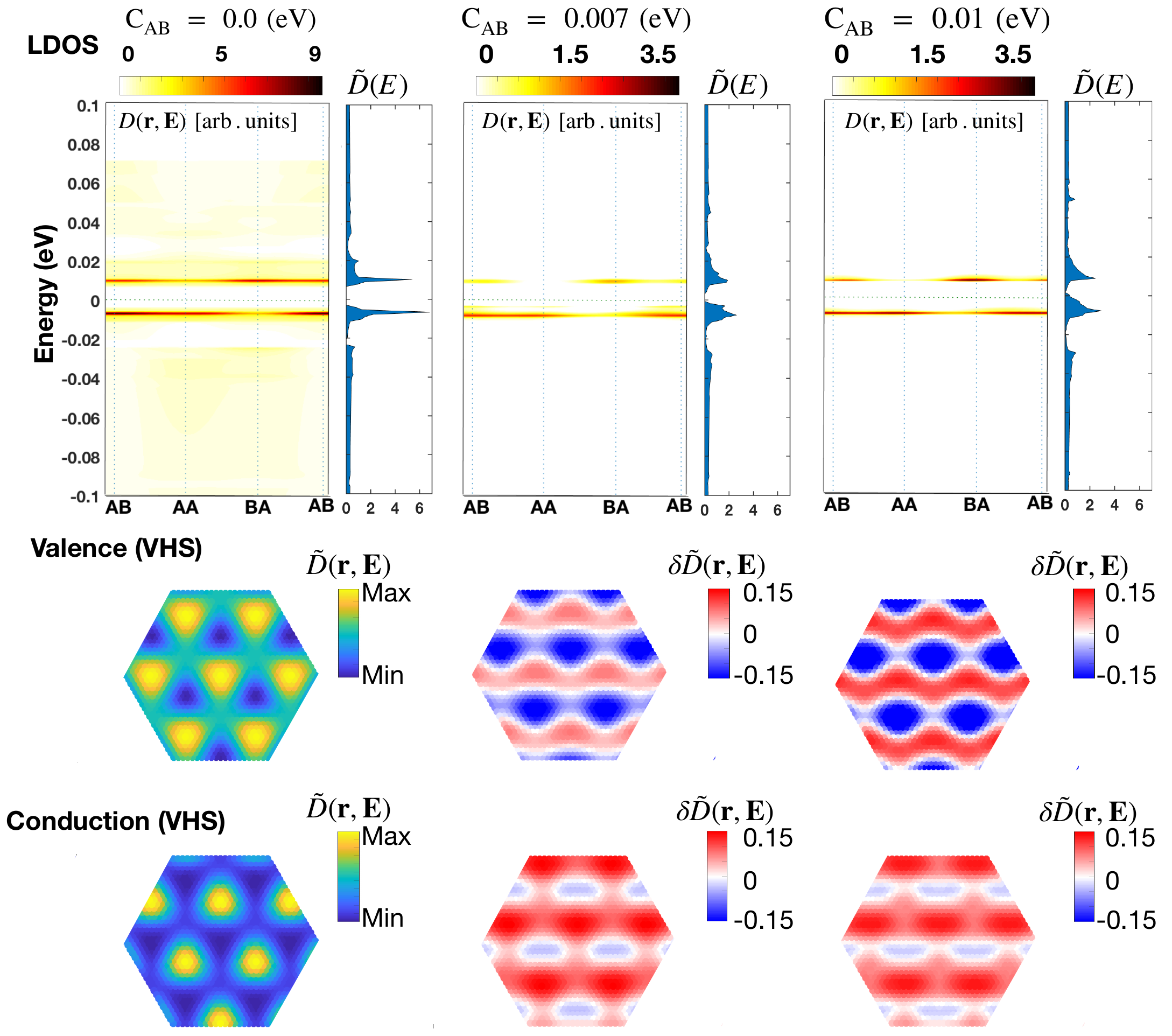} 
\caption{(color online) 
The local density of states (LDOS, $D({\vec{r}}, E)$) along with the total density of states (DOS, $D(E)$ ) projected vertically to account the LDOS with respect to the inclusion of off-diagonal term in the moire Hamiltonian in TLG/BN. For $\xi =1$ with $\Delta = 0.01$ eV, the LDOS and DOS are obtained for different off-diagonal terms ($C_{AB} = 0, 0.007, 0.01$ eV), that are discussed in Fig.~\ref{elect}. The real space representation of the normalized LDOS ($\tilde{D}({\vec{r}}, E)$) is presented for each Van Hove singularity (vHS) at the valence and conduction bands  $C_{AB} = 0$, 
and for the cases when $C_{AB} \ne 0$ we calculate their differences $\delta\tilde{D}({\vec{r}}, E) = \tilde{D}_{C_{AB} \ne 0}({\vec{r}}, E) -\tilde{D}_{C_{AB} = 0}({\vec{r}}, E)$. The normalized LDOS is defined as $\tilde{D}({\vec{r}}, E) = D({\vec{r}}, E)$/max$[D({\vec{r}}, E)]$ and we have plotted them at the van Hove singularity energies $E = E_{vHS}$.}
\label{ldos}
\end{figure}

\begin{figure*}[t]
\label{schematic}
\includegraphics[width=18cm,angle=0]{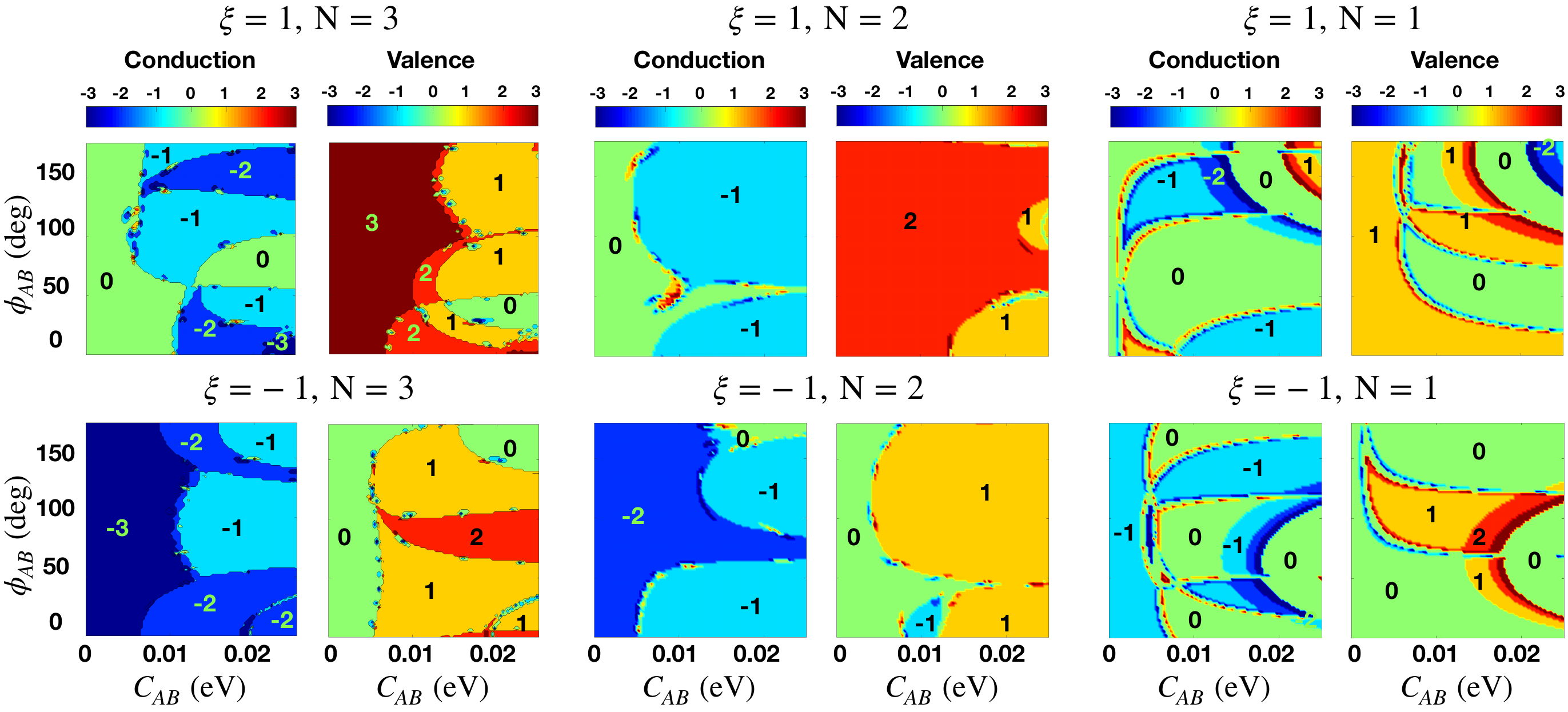} 
\caption{(color online) 
The valley Chern number phase diagrams for the low energy conduction and valence bands of $N$-layer graphene boron nitride moire superlattices in the parameter space that defines the strength and shape of off-diagonal terms 
through the parameters ($C_{AB}$ ($0 \sim 0.025$)~eV and $\phi_{AB}$ ($0^{\circ} \sim 180^{\circ}$)) in the moire Hamiltonian. 
The valley Chern number phase diagrams are obtained for $\xi =1$ (top panels) and $\xi = -1$ (bottom panels) 
with $N= 3$ (left), $N = 2$ (middle) and $N = 1$ (right).}
\label{fig:Chern}
\end{figure*}

\section{Results and Discussions}
Here we present the electronic structure of ABC multilayer graphene on hBN
under the effect of an off-diagonal moire pattern term in Eq.~(\ref{vecpot})
by allowing variations of the parameters between $C_{AB} = 0 \sim 0.025$~eV and $\phi_{AB} = 0 \sim \pi$ for $N=1,2,3$ systems.
We have verified that changing the diagonal moire parameters alone in a similar range of magnitude and phase
did not modify the valley Chern numbers of the low energy bands which are 0 or $N$. 
In the following we assess the impact of this term in particular in the valley Chern number of the low energy bands 
for fixed diagonal moire patterns and discuss the 
real-space anisotropies introduced in the local density of states.

\subsection{Topological phase transitions in the low energy bands}\label{sec:Topology}
The main finding in this work is that the introduction of an off-diagonal interlayer coherence moire
patterm term through a finite $C_{AB} \neq 0$ can switch the valley Chern number of the low energy bands
in the system as we illustrate in Fig.~\ref{elect},
where the vector moire potential term $H_{\xi}^{A} (\vec{d})$ defines a pseudospin field on the graphene layer contacting hBN.

 The inclusion of off-diagonal terms causes significant changes in the electronic structure and triggers 
 topological phase transitions, as shown in Fig.~\ref{elect}(a). 
The band structures for $N = 3$ (TLG/BN) were obtained with $\xi = 1$ with a constant value of interlayer potential difference $\Delta = 0.01$~eV. 
The continuous variation of the $C_{AB}$ term with $\phi_{AB} = 0$ leads to a gap closure between
the low energy valence band and the higher energy bands between $\tilde{K}$ and $\tilde{K^{\prime}}$ for a value of $C_{AB} = $ 0.007 eV, and the gap reopens for larger values of  $C_{AB}$ as shown in Fig.~\ref{elect}(a). Under the electric field, the gap at the primary Dirac point and the avoided gaps at the moire mini-Brillouin-zone (mBZ) boundaries isolate the low-energy bands near charge neutrality.

Introduction of a finite $C_{AB}$ parameter also breaks the triangular rotational symmetry of the moire Brillouin 
zone that can be visualized in the Berry curvature distribution in Fig.~\ref{elect}(b). 
The Berry curvatures of the low energy $n^{th}$ low-energy bands are calculated through~\cite{rmp_berry} 
\begin{equation}
    \Omega_n (\vec{k}) = -2 \sum_{n' \neq n} Im \left[  \langle u_n | \frac{\partial H}{ \partial k_x} | u_{n'} \rangle    \langle u_{n'} |  \frac{\partial H}{ \partial k_y} | u_{n} \rangle / \left( E_{n'} - E_n \right)^2  \right],
\end{equation}
\noindent where, for every $k$-point, we obtain sums through all the neighboring $n'$ bands; $| u_n \rangle$ are the moire superlattice Bloch states, and $E_n$ are the eigenvalues. Based on the Berry curvature, the valley Chern number of the $n^{th}$ band is obtained from  $C = \int {\rm d^2} \vec{k} \,\, \Omega_{n}(\vec{k})/(2\pi)$ integrated in the moire Brillouin zone. The Berry curvatures for the valence band for three values of $C_{AB}$ (before, after and at the transition) are compared in Fig.~\ref{elect}(b). 
Band closure between the low energy and higher energy bands is observed between  
moire Brillouin zone corners
$\tilde K^{\prime}_{1} = 2\pi(-\frac{2}{3}, \frac{1}{\sqrt{3}})$ and $\tilde K_{1} = 2\pi(\frac{1}{3}, \frac{1}{\sqrt{3}})$, 
and is absent between another set of $\tilde K^{\prime}_{2} =  2\pi(\frac{2}{3}, 0)$ and 
$\tilde K_{2} = 2\pi(\frac{1}{3}, \frac{1}{\sqrt{3}})$ in the mBZ.
This asymmetry between the initially equivalent $\tilde{K}_1$ and $\tilde{K}_2$ or
$\tilde{K}^{\prime}_1$ and $\tilde{K}^{\prime}_2$ mini-valleys 
indicates the rotational symmetry breaking introduced by the $C_{AB}$ term. 

The band closure as a function of $C_{AB}$ indicates a possible topological transition of the bands. 
Indeed, the valley Chern number $C = 3$ with $C_{AB} = 0$ has changed to $C = 2$ after the band closure
and re-opening. 
We present the low energy bands valley Chern numbers in Fig.~\ref{elect}(c), 
where the valley Chern number very near the band closure point ($C_{AB}$ = 0.007 eV) is $C = 0$. 
The conduction band valley Chern number undergoes a transition from  $C = 0$ to a non-zero valley Chern number $C = -2$ as shown in Fig.~\ref{elect} (c).

For $\xi = 1$ hBN alignment and $N=2$ bilayer graphene (BG/BN) where we allow $C_{AB}$ to change while keeping $\phi_{AB}$ = 0 
we see that the valence band exhibits a topological phase transition from $C = 2$ to $C = 1$ as shown in Fig.~\ref{elect}(d).
However, for $N =1$ monolayer graphene aligned with hBN (G/BN)
the valence band retains the same valley Chern number of $C = 1$ even with increasing values of  $C_{AB}$. 

\subsection{Broken rotational symmetry nematic local density of states}\label{sec:ldos}
We had shown in Fig.~\ref{elect} for a particular case of TLG/BN with $\xi = 1$ and $\Delta = 0.01$~eV 
that the addition of an off-diagonal term in the moire Hamiltonian can trigger a topological phase transition.
Here we report calculations of local density of states (LDOS) in order to distinguish how the
added off-diagonal terms can modify the LDOS profile $D(\vec{r}, {E})$. 
In Fig.~\ref{ldos} we show the LDOS and density of states (DOS) for the low energy valence and conduction 
bands for theree values of $C_{AB}$, namely 0, 0.007, and 0.01~eV. 
The van Hove singularities in the DOS near the charge neutrality indicate the presence of flat bands.
The conduction band has a localization at BA stacking, whereas the valence band has localization at AA stacking. 
With a finite off diagonal term $C_{AB} \ne 0$ the vHS peak is broadened slightly
but the localization remains at the same stacking as illustrated in the LDOS plots.
We use the normalized LDOS defined as $\tilde{D}(\vec{r}, E) = D(\vec{r}, E)$/max$[D(\vec{r}, E)]$ in our plots
and find the influence of off-diagonal moire $C_{AB}$ term on the LDOS map through 
the difference $\delta\tilde{D}(\vec{r}, E) = \tilde{D}_{C_{AB} \ne 0}(\vec{r}, E) -\tilde{D}_{C_{AB} = 0}(\vec{r}, E)$.  
This quantity shows how the inclusion of a finite off-diagonal term leads to anisotropic LDOS profiles and breaks the triangular rotational symmetry of the solutions.

\subsection{Pattern shape dependent Chern number phase diagrams}\label{sec:Chern_Phase}
The moire pattern shapes that mix inversion symmetric and asymmetric components 
can be calibrated through the phase parameter $\phi$ that in turn controls the shapes of the triangular moire patterns
in the first harmonic expansion, see the appendix for more details. 
In Fig.~\ref{elect} we had noted the change in the valley Chern numbers with $C_{AB} \neq 0$ and $\phi_{AB} =0$ combinations.
However, there could be additional valley Chern number phases when $\phi_{AB}$ is allowed to take a finite value. 
Indeed, we find this is the case and we have presented in Fig.~\ref{fig:Chern} the valley Chern number 
phase diagram in the parameter space of $C_{AB} = 0 \sim 0.025$~eV and $\phi_{AB} = 0 \sim \pi$ for  TLG/BN, 
BG/BN and G/BN as concrete examples of $N$-chiral systems up to three layers.

The results markedly depend on the orientation of the BN layer and 
for $\xi = 1$ we generally obtain nonzero valley Chern number equal to $C = N$ 
while the conduction band is trivial with $C = 0$ for $N = 1, 2, 3$. 
When we change the orientation of the substrate moire pattern by setting $\xi = -1$ the 
conduction band becomes non-trivial acquiring an opposite sign valley Chern number $C = -N$,
while the valence band acquires $C = 0$. 
It is noted that the valley Chern number of the valence/conduction bands are sensitive to both 
the off-diagonal parameters $C_{AB} \ne 0$ and $\phi_{AB} \ne 0$ provided that the former is
sufficiently large and allows for additional valley Chern number transitions for both initially trivial and non-trivial bands.

In the case of TLG/BN, $N = 3$ with $\xi = 1$, the valley Chern number of the valence (conduction) 
band is $C = 3 \, (0)$ for $C_{AB}$ = 0. 
The valley Chern number of the valence (conduction) band changed to $C = 2 \, (-2)$ for sufficiently large $C_{AB}$ with $\phi_{AB}$ = 0. 
For $\xi = -1$, the valley Chern number of valence (conduction) band is $C = 0 \, (-3)$ for $C_{AB}$ = 0, 
and changed to $C = 1 \, (-2)$ with increasing value of $C_{AB}$. 

Similarly,  for BG/BN, $N = 2$ with $\xi = 1$ the valence (conduction) band has a phase transition from 
$C = 2 \, (0)$ to $C = 1\, (-1)$. 
However, with $\xi = -1$, the valence/conduction band shows a transition from $C = 0\, (-2)$ 
to $C = -1\, (-1)$. 

For G/BN, $N = 1$ with $\xi = 1$, the valence/conduction band has a valley Chern number $C = 1\,(0)$.
Even though the valley Chern number of the valence band did not change with increasing $C_{AB}$, 
the conduction band valley Chern number changed to $C = 1$. 
For the case of $\xi = -1$, the valence band valley Chern number remains unchanged
while it does change for the conduction band when we modify $C_{AB}$.

\section{Summary and discussion}\label{sec:conclu}
We have explored the phase diagram map for the valley Chern numbers of the low energy valence and conduction bands
of rhombohedral $N=1,2,3$ layer graphene boron-nitride superlattices for different moire patterns. 
The intra-sublattice diagonal moire patterns produces low energy bands whose valley Chern number
magnitudes are zero or proportional to layer number $N$. 
The absolute value of the maximum valley Chern number followed the number of graphene layers $N = 1, 2, 3$ for all 
possible moire patterns within the first harmonic approximation. 
For zero off-diagonal patterns we find that the valence band has a valley Chern number equal to the number of layers $C = N$ when $\xi = 1$, while the conduction band is $C = -N$ when $\xi = -1$.

However, introduction of off-diagonal interlayer coherence moire pattern terms captured
through the magnitude $C_{AB}$ and phase $\phi_{AB}$ parameters allows to trigger topological phase transitions 
giving rise to valley Chern numbers that are smaller than the number of layers,
a behavior that can be traced mainly to the variations in the electron-hole secondary Chern weights near 
the moire Brillouin zone boundaries. 
Thus, experimentally observed quantum anomalous Hall effects in TLG/BN~\cite{Nat_TLGBN} 
compatible with $C=2$ rather than the expected $C= 3$ from layer number
could result from rotational symmetry breaking interlayer coherence terms introduced by the Coulomb interactions. 
Our model off-diagonal contributions could also result from higher order corrections of the moire potentials mediated by interlayer tunneling. 
Broken rotational symmetry in the mBZ upon inclusion of off-diagonal terms in the Hamiltonian
is evidenced from the Berry curvature distribution in momentum space and the LDOS maps. 

In summary, we have presented valley Chern number phase diagrams in the 
parameter space that defines the off-diagonal moire pattern of the model Hamiltonian of rhombohedral 
$N$-chiral multilayer graphene subject to moire scalar and vector potentials.
The vector potential moire patterns which are often ignored in the low energy Hamiltonian models of
 $N$-chiral multilayer graphene BN superlattices turned out to play a more prominent role than the 
scalar potentials for triggering topological transitions to phases with valley Chern numbers that are
different to those dictated by layer number. 
Our results points to the possibility of triggering topological phase transitions by breaking the 
triangular rotational symmetry through pseudomagnetic field vector potentials generated by 
moire strain patterns.
These can result for example from real strain fields whose bond distortions give rise to unequal electron hopping 
probabilities to the neighboring atoms, or from virtual effective strains 
due to high order interlayer electron hopping processes or Coulomb interactions.

\begin{acknowledgments}
D.A.G.G. acknowledges partial support from the Universidad de Antioquia, Colombia, under initiative CODI ES84180154 
\textit{Estrategia de sostenibilidad del Grupo de F\'{\i}sica At\'omica y Molecular} and projects CODI-251594 and 2019-24770.
Additionally, the authors are grateful to Professor Dr. Jorge Mahecha from the University of Antioquia for his advice and assistance in this work. 
We acknowledge financial support from the National Research Foundation of Korea (NRF)
through grants NRF-2020R1A2C3009142 for B.L.C., NRF-2020R1A5A1016518 for Y. P., 
the Zhejiang Provincial Natural Science Foundation of China (Grant No. LY19A040003) for J. H. S.,
and the Basic study and Urban convergence R\&D research fund of the University of Seoul (2019) for J. J.
We acknowledge computational support from KISTI through grant KSC-2020-CRE-0072.

\end{acknowledgments}

\bibliographystyle{apsrev4-1}
\bibliography{myref}

\newpage

%\begin{widetext}
\begin{appendix}
\section{Inversion symmetric and asymmetric moire patterns}
\begin{figure}[!htb]
\centering
\includegraphics[width=9cm]{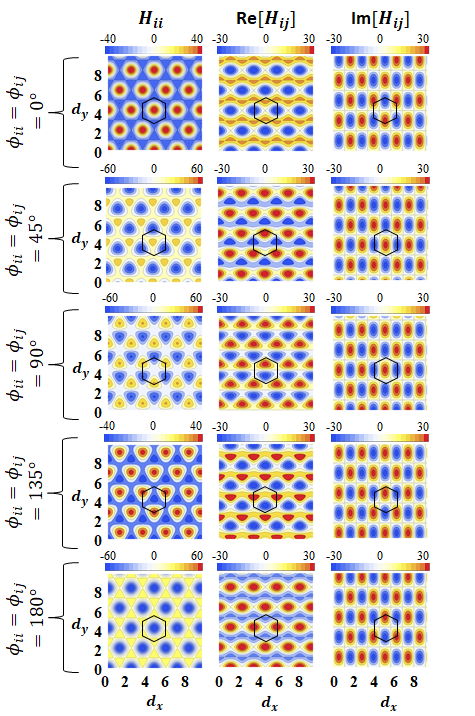}
\caption{Intralayer Hamiltonian elements as a function of sliding vector $\vec{d}$ for different phase angles $\varphi_{ii}$ and $\varphi_{ij}$. The black hexagon represents the real space presentation of moire supercell. 
In the real space the inversion symmetric potential for the diagonal term $H_{ii}$ is associated 
with $\varphi_{ii} = n \pi$ but the inversion asymmetric potential for the off-diagonal term 
$H_{ij}$ is associated with $\varphi_{ij}= (2n+1) \pi / 2$ where $N$ is an integer.}
\label{fig:intra}
\end{figure}

Here we discuss in more detail the breakdown of the moire patterns into 
inversion symmetric and asymmetric terms~\cite{Wallbank2013}
for the diagonal scalar moire pattern in Eq.~(\ref{diagterm}) and  
off-diagonal vector potential-like term in Eq.~(\ref{vecpot}).
Illustrations of the diagonal and off-diagonal moire patterns for different phase angles
are in Fig.~\ref{fig:intra}.

\begin{figure}
\includegraphics[width=8cm,angle=0]{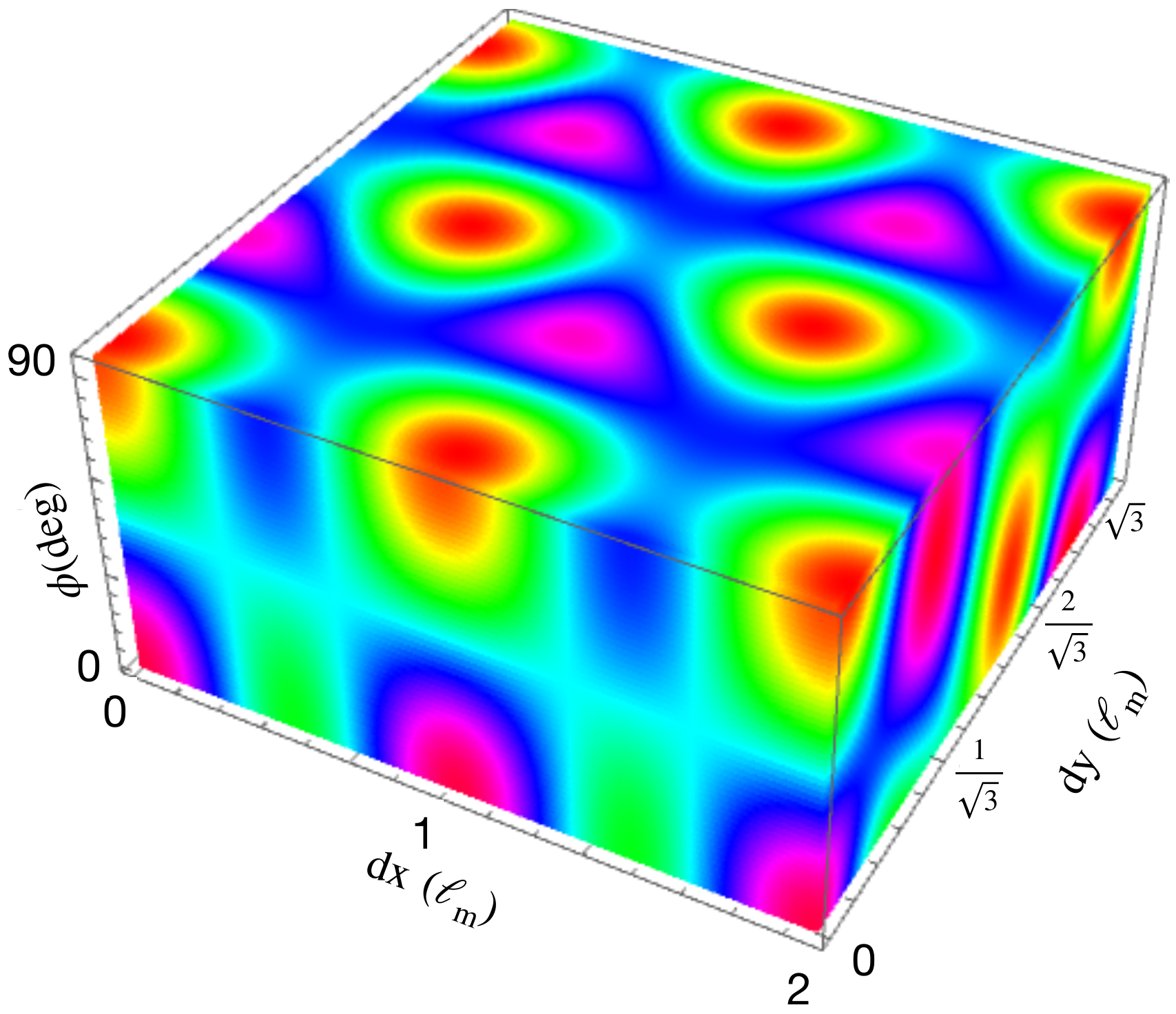} 
\caption{Dependence of the scalar moire potential term 
$M(\vec{r})= 2C \, {\rm Re}\left[ f(\vec{r}) e^{i\phi}  \right]$ on 
the local stacking vector $\vec{d}$ and the phase angle $\phi$. 
The local stacking coordinate vector and real space position is related through 
$\vec{d}(\vec{r}) = \varepsilon \vec{r} + \theta \hat{z} \times \vec{r}$ in the small angle approximation. 
The moire pattern in Eq.~(\ref{symasym}) will be symmetric for $\phi_{ii} = n \pi$ and
asymmetric when $\phi_{ii} = (2n + 1) \pi/2 $ for integer $n$ values,
and a combination of both for intermediate phase angles.
 }
\label{fig:M}
\end{figure}

%\subsection{Diagonal term}
%
The diagonal element of the moire potential in Eq.~(\ref{diagterm}) involving 
a scalar function $V_{AA/BB}^M(\vec{r})$ is defined in terms of 
$C_{AA/BB} = C_{ii},~\phi_{AA/BB} = \phi_{ii}$ and for $\xi = 1$ 
we have
\begin{eqnarray} 
M_{ii}(\vec{r})&=& \label{symasym}
2C_{ii}  \, {\rm Re}\left[ f(\vec{r}) e^{i\phi_{ii}}  \right] \\ \nonumber
&=&C_{ii}\left( f_1(\vec{r}) \cos{\phi_{ii}} + f_2(\vec{r}) \sin{\phi_{ii}} \right), 
\end{eqnarray} 
with $ f_1(\vec{r})=\sum_{m=1}^6 e^{i {\tilde{G}}_m \vec{r}}$ (symmetric function)
and $f_2(\vec{r})=i\sum_{m=1}^6(-1)^{m-1} e^{i {\tilde{G}}_m \vec{r}}$ (antisymmetric function),
where we use six moire reciprocal lattices $\tilde G_{m=1..6} = \hat R_{2\pi(m-1)/3} {\tilde G_1}$ 
successively rotated by 60$^{\circ}$ as introduced in the main text. 
Hence, the moire pattern in Eq.~(\ref{symasym}) will be symmetric for $\phi_{ii} = n \pi$ and
asymmetric when $\phi_{ii} = (2n + 1) \pi/2 $ for integer $N$ values,
and a combination of both for intermediate phase angles.
In Fig.~\ref{fig:M}, we illustrate the $M(\vec{r})$ patterns in real space for different values of the angle $\phi_{ii}$.  
%
%

\begin{figure*}[hb]
\includegraphics[width=1.0\linewidth]{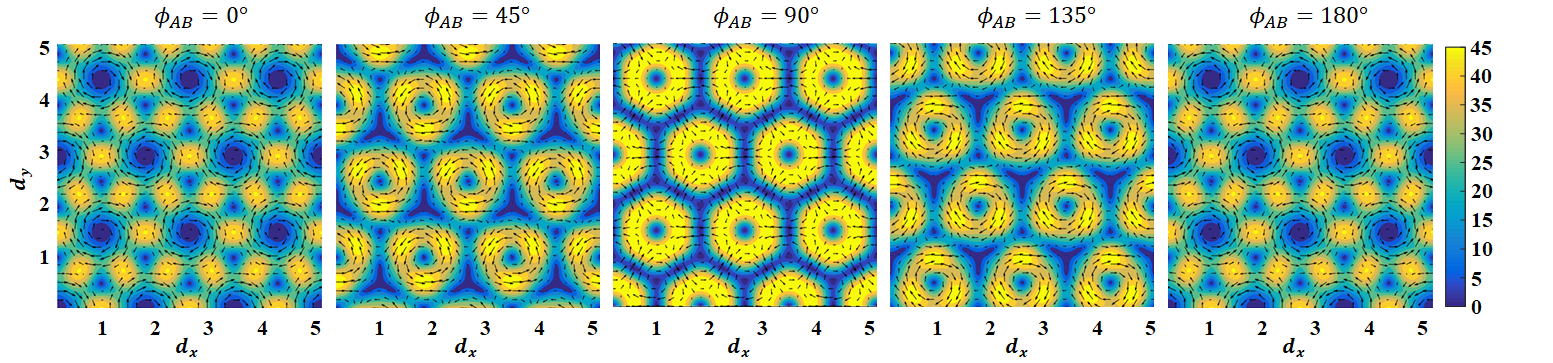} 
\caption{Dependence of the vector moire potential $\vec{A}(\vec{r})$ on phase angle $\phi_{AB}$ that decide the relative presence of inversion symmetric and asymmetric pattern components as in Fig.~\ref{fig:M} for the diagonal terms. The real and imaginary parts of the inter-sub lattice tunneling sets the direction of the pseudo magnetic vector potential (arrows) which strongly depends on the phase angle $\phi_{AB}$.}
\label{fig:Ar}
\end{figure*}

%\subsection{Off-diagonal term}
%
The off-diagonal $H_{ij}(\vec{r})$ term with $i \neq j$ given in Eq.~(\ref{vecpot}) has a
vector potential term $\vec{A}(\vec{r})$ defined in Eq.~(\ref{vecpotarr}) that can 
be represented by its magnitude and orientation as in Fig.~\ref{fig:Ar},
\begin{eqnarray} 
A^\xi(\vec{r}) &=&  V^M_{AB}\vec\nabla_{\vec{r}} {\rm Re}\left[e^{i\phi_{AB}}f(\vec{r})\right]  \label{vecpotarr}
\\ \nonumber
&=&V^M_{AB}\vec\nabla_{\vec{r}} {\rm Re}\left[\sum_{m=1}^{6}e^{i\xi{ {\tilde G_m}}\cdot{\vec r}+\phi_{AB}}\left(1+\frac{ \left( 1+(-1)^m \right)}{2}\right) \right].
% \nonumber
%&=&V^M_{AB}\vec\nabla_{\vec{r}} \sum_{m=2,4,6} {\rm cos} \left({\tilde G_m}\cdot{\vec r}+\phi_{AB}\right) \\ \nonumber
%%&=&V^M_{AB}\vec\nabla_{\vec{r}} \sum_{m=2,4,6} {\rm cos} \left({\tilde G_{mx}}x + {\tilde G_{my}}y+\phi_{AB}\right) \\  \nonumber
%&=&V^M_{AB} \sum_{m=2,4,6} {\tilde G_m}\cdot {\rm sin} \left({\tilde G_m}\cdot{\vec r}+\phi_{AB}\right) \\
%%&=&V^M_{AB}\sum_{m=2,4,6} {\rm sin} \left({\tilde G_m}\cdot{\vec r}+\phi_{AB}\right)\left\{{\tilde G_{mx}}, {\tilde G_{my}}, 0\right\}   
\end{eqnarray}

%term is altered with respect to $\phi_{ij}$ phase angles.  
%In Fig.~\ref{fig:Ar}, we show the variation of  $\vec{A}(\vec{r})$ for different $\phi_{AB}$ phase angles. 

%We present the moire potential matrix elements as a function of the local stacking vector $\vec{d}$ through $H_{ii}(K,\vec{d}(\vec{r})) =  V_{AA/BB}^M(\vec{r})$ and $H_{ij}(K,\vec{d}(\vec{r})) = A^\xi(\vec{r})$ where $H_{ii}$ are site energies and $H_{ij}$ are the inter sublattice tunneling within the layers ($i$ and $j$ represent $A$ and $B$). The local stacking vector and the real space position is mutually related through $\vec{d} = \varepsilon \vec{r} + \theta \vec z \times \vec{r}$ in the small angle approximation. 
%

\end{appendix}
%\end{widetext}
\end{document}